\documentstyle[12pt]{article}

\def\bc{\begin{center}}
\def\ec{\end{center}}
\def\be{\begin{equation}}
\def\ee{\end{equation}}
\def\myappendix{\par
 \setcounter{section}{0}
 \setcounter{subsection}{0}
 \setcounter{equation}{0}
 \setcounter{table}{0}
 \def\appendixname{Appendix}
 \def\appesection{\setcounter{equation}{0}\section}
 \def\@thesection{\Alph{section}}
 \def\thesection{\appendixname\hskip 1.10ex\Alph{section}}
 \def\thesubsection{\@thesection.\arabic{subsection}}
 \def\theequation{\@thesection.\arabic{equation}}
 \def\thetable{\@thesection.\arabic{table}}}

\newcommand{\labar}{\overline{\Lambda}}
\newcommand{\lb}{\overline{\Lambda}}
\newcommand{\lqcd}{\Lambda_{\mathrm QCD}}

\newcommand{\beq}{\begin{equation}}
\newcommand{\eeq}{\end{equation}}
\newcommand{\beqn}{\begin{eqnarray}}
\newcommand{\eeqn}{\end{eqnarray}}

\def\vdir{v\kern-7.8pt\Big{/}}
\def\pdir{p\kern-7.8pt\Big{/}}
 
\newcommand{\deltambar}{\delta\overline{m}}

\begin{document}
\pagestyle{empty} 
\vspace{-0.6in}
\begin{flushright}
CERN-TH/96-163 \\
ROME prep. 96/1144 \\
SHEP 96-14\\ 
FTUV 96/21 - IFIC 96/23
\end{flushright}
\vskip 0.4 cm
\centerline{\LARGE{\bf{A High Statistics Lattice Calculation of}}}
\vskip 0.2cm
\centerline{\LARGE{\bf{The B-meson Binding Energy.}}}
\vskip 0.4cm
\centerline{\bf{V. Gim\'enez$^{1}$,
G. Martinelli$^{2}$ and C. T. Sachrajda$^{3,*}$}}
\centerline{$^1$ Dep. de Fisica Teorica and IFIC, Univ. de Valencia,}
\centerline{Dr. Moliner 50, E-46100, Burjassot, Valencia, Spain.} 
\centerline{$^2$ Dip. di Fisica, Univ. ``La Sapienza"  and}
\centerline{INFN, Sezione di Roma, P.le A. Moro, I-00185 Rome, Italy.}
\centerline{$^3$ Theory Division, CERN, 1211 Geneva 23, Switzerland.}
\abstract{
We present a high statistics lattice calculation of the B-meson binding energy 
$\labar$
of the heavy-quark inside 
the pseudoscalar  B-meson. Our numerical results have been obtained from 
several  independent numerical
simulations at $\beta=6.0$, $6.2$ and $6.4$, and using,  for
the meson correlators, the results obtained by the
APE group at the same values of $\beta$. Our best estimate, obtained
by combining results at different values of $\beta$, is 
$\labar = 180^{+30}_{-20}$~MeV.
For the $\overline{MS}$ running mass,
we obtain $\overline {m}_b(\overline {m}_b)
=4.15 \pm 0.05 \pm 0.20$~GeV, in reasonable agreement with
previous determinations. The systematic error is
the truncation of the perturbative series in the  matching condition
of the relevant operator of   the Heavy Quark Effective Theory.}
\vskip 0.5 cm
\begin{flushleft}
CERN-TH/96-163\\
July 1996 \\
\end{flushleft}\vskip 1.5cm
\centerline{$^*$ On leave of absence  from  Dept. of Physics,  
University of Southampton, 
 Southampton SO17 1BJ,UK.} 
\vfill\eject
\pagestyle{empty}\clearpage
\setcounter{page}{1}
\pagestyle{plain}
\newpage 
\pagestyle{plain} \setcounter{page}{1}
 
\section{Introduction} \label{intro}

In a previous paper \cite{cgms} we have presented the first results
for the binding and kinetic energies of the $b$--quark in a $B$-meson,
obtained by using the lattice version of the Heavy Quark Effective
Theory (HQET) \cite{eihil}. These are the non-perturbative parameters
$\labar$ and $\lambda_1$, which control the $O(1/m_b)$ and
$O(1/m_b^2)$ corrections to the spectroscopy and inclusive decays of
$B$--mesons, and which also appear in the theoretical predictions for
several exclusive processes~\footnote{In addition there is the
  parameter $\lambda_2$, which is the matrix element of the
  chromomagnetic operator. This parameter gives the leading term in
  the hyperfine splitting ($m_{B^*}-m_B$), and so can be estimated
  directly from the experimental data.}. $m_b$, $m_B$ and $m_{B^*}$
denote the masses of the $b$-quark and of the $B$ and $B^*$ mesons
respectively. In this letter we report on a detailed study of
$\labar$, which improves on that in ref.~\cite{cgms} in two ways:
\begin{enumerate}
\item[i)] We have collected a much larger set of gauge field
  configurations than was used in ref.~\cite{cgms}, with a consequent
  reduction of the statistical error in the determination of $\labar$.
  Moreover, the more accurate knowledge of the heavy quark propagator
  has allowed for an improved study of its infrared behaviour (i.e.
  its behaviour at large time separations). This will be discussed in
  detail in the following.
\item[ii)] We also present new results at a value of the lattice
  spacing which is smaller than those considered in ref.~\cite{cgms},
  corresponding to $\beta=6.4$. Combining the results at three
  different values of $\beta$ ($\beta = 6.0$, $6.2$ and $6.4$) it is
  possible to study the dependence of $\labar$ on the lattice spacing.
  Within the precision of our computations, we find that the results
  are independent of the lattice spacing.
\end{enumerate}

The presence of renormalon singularities in the pole mass
\cite{bigi,beneke} implies that the most intuitive definition of the
parameter $\labar$, i.e. $\labar\equiv m_B-m_b^{\mathrm pole}$, where
$m_b^{\mathrm pole}$ denotes the pole mass of the $b$--quark, does not
correspond to a physical quantity. Thus if $\labar$ is to be
introduced into phenomenological applications (which is not necessary,
see below), alternative definitions have to be given. For example, it
is possible to define $\labar$ from the experimentally measured value
of some physical quantity, for which the theoretical prediction
depends on $\labar$. Since $\labar$ does not contribute directly to
the leading behaviour of any physical quantity, always appearing as an
$O(1/m_b)$ correction, in order to extract $\labar$, it is first
necessary to subtract the perturbation series for the leading term
(for a detailed explanation see ref.~\cite{emphasis}). This series has
a renormalon ambiguity, implying that its sum is not uniquely defined,
but the subtraction is performed up to some finite order of
perturbation theory. The physical quantity being used, and the order
of the perturbation series which has been subtracted, constitute the
definition of $\labar$\ ~\footnote{ Notice that such definitions of 
$\labar$ retain some dependence on the mass of the heavy quark.}.

Lattice simulations provide the opportunity to compute $\labar$,
defined in some suitable way, non-perturbatively. In ref.~\cite{ms} a
definition of $\labar$ was proposed which is free of renormalon
ambiguities, and from which the linear ultraviolet divergence (i.e.
the divergence of $O(a^{-1})$, where $a$ is the lattice spacing) has
been subtracted non-perturbatively.  The necessity for
non-perturbative subtractions of power divergences has been stressed
in ref.~\cite{mms}. This definition of $\labar$, which will be
explained in section~\ref{labardef}, is obtained after imposing a
non-perturbative renormalisation condition on the heavy quark
propagator in the Landau gauge. Even though, in practice, the value of
$\labar$ defined in this way is obtained using lattice simulations, it
is in fact independent of the method of regularisation.

Although it may be convenient to try to introduce a parameter $\lb$,
which is of $O(\lqcd)$ and is free of renormalon ambiguities, it is
not necessary to do so. One can either relate two or more physical
quantities directly to the required precision~\footnote{This is done
  implicitly when $\labar$ is defined using one physical quantity, and
  used in the prediction for a second one.}, or determine physical
quantities from non-perturbative computations, such as lattice
simulations~\cite{emphasis}.  Below, as an example of the computation
of a physical quantity, we present the results for the $b$-quark mass
($\overline {m}_b$) renormalized in the $\overline{MS}$ scheme at the
scale $\mu=\overline {m}_b$.  The main uncertainty in this quantity
comes from the truncation at $O(\alpha_s)$ of the perturbative factor
necessary to match the quark mass computed in the HQET to $\overline
{m}_b$. We briefly discuss how to reduce this error which we currently
estimate to be about $200$ MeV \cite{emphasis}.
\par Any definition of $\labar$ which is of $O(\lqcd)$ and is independent
of any renormalization scale, necessarily involves Green functions
defined at large distances, Although we have introduced such a definition
in ref.~\cite{ms}, which is free of renormalon singularities,
we cannot prove that non-perturbative confining effects do not 
invalidate the proposed procedure. An important aspect of this paper
is to study the infrared behaviour of the heavy quark propagator
in order to search for such effects.  They are not visible within the
precision of our  lattice calculations, up to distances of about
1.5 fm, which is the largest distance which we can reach.
\par
The plan of the paper is the following. In sec.~\ref{labardef} we
recall the relevant formulae for our definition of $\lb$~\cite{ms}; in
sec.~\ref{numerical} we describe the numerical calculation of $\labar$
and $\overline{m}_b$ and discuss the main results of this study; in
the conclusions we present the outlook for future developments and
applications of the method discussed in this paper.

\section{Non-perturbative definition of $\labar$} 
\label{labardef}

In this section we define the non-perturbative renormalisation
prescription which we will use to calculate the ``physical" value of
$\labar$ \cite{ms}. A detailed discussion of our procedure has
been presented in our previous papers \cite{cgms,ms}, and so we only
exhibit here those formulae which are necessary to the understanding
of the numerical results.

Consider the following Lagrangian of the lattice formulation of the HQET:
\be
{\cal L}_{\mathrm eff} \,=\, \frac{1}{1+ a \,\delta m}
\Bigl( \bar h(x)\, D_4\, h(x)\, +\, 
\delta m\, \bar{h}(x)\, h(x)\Bigr),
\label{eq:l0effp}
\ee where the factor $1/(1+a \, \delta m)$ has been introduced to
ensure the correct normalization of the heavy quark field $h$, and
the covariant derivative is defined by
\be
D_4\,f(t)=1/a\,\left( f(t)-U^{\dagger}_4(t-a)f(t-a) \right)\ .
\label{eq:d4def}\ee
Lattice computations are frequently performed by choosing the
bare-mass ($\delta m$) to be zero.  We propose instead to choose a
value of the bare mass by imposing a renormalisation condition on the
heavy quark propagator. With our prescription, the heavy-quark
propagator can be made finite, up to the standard logarithmic
divergences associated with the renormalization of the heavy-quark
field.

\par The mass counter-term  can be chosen in many different ways.
One possibility is to define it by studying the behaviour of
heavy-quark propagator $S(\vec x ,t)$ in some gauge (we will
use the Landau gauge) at large values of the time:
\beqn 
-\, \delta \overline{m}\, \equiv\, \frac{\ln(1+a \,\delta m )}{a}\,
=\, \lim_{t \to \infty} \frac{1}{a}\, \ln\left[\frac{{\mathrm
      Tr}\Bigl(S(\vec x ,t+a)\Bigr) } {{\mathrm Tr}\Bigl(S(\vec x
    ,t)\Bigr)} \right]\ , 
\label{ct} \eeqn
where the traces are over the colour quantum numbers, and we have
assumed that the large time limit of the ratio in eq.~(\ref{ct})
exists. Our numerical results, to be discussed in detail in the next
section, support the validity of this assumption and the use of
eq.~(\ref{ct}) is our preferred definition of $\deltambar$. In words,
(\ref{ct}) is the condition that the subtracted heavy quark propagator has no
exponential fall (or growth) at large times. If, instead, the bare
mass $\delta m$ is chosen to be zero, loop corrections generate a mass
of $O(1/a)$, which leads to an exponential behaviour in time.

\par A possible alternative definition is given by
\beqn 
-\, \delta \overline{m}(t^*)\, \equiv\, \frac{1}{a}\, \ln
\left[\frac{{\mathrm Tr}\Bigl(S(\vec x ,t^*+a)\Bigr)} {{\mathrm
      Tr}\Bigl(S(\vec x ,t^*)\Bigr)}\right] , 
\label{ct1} \eeqn 
for some chosen time $t^*$. The corresponding definition of $\labar$,
see eq.~(\ref{lare}) below, will clearly depend on the choice of
$t^*$: $t^*$ parametrizes the renormalisation prescription dependence
and can be considered as the renormalisation point in coordinate
space.  For physical matrix elements the dependence on $t^*$ is
eliminated in the matching of the QCD action and operators with those
of the HQET.  The use of the propagator at small times,
$t^*\Lambda_{QCD}\ll 1$, to define $\delta \overline{ m}(t^*)$, and
hence $\lb(t^*)$, does not require any assumption about the large time
behaviour of the heavy quark propagator and in section~\ref{numerical}
we will also present the results for $\lb(t^*)$ obtained in this way.
However $\lb(t^*)$ contains terms of $O(1/t^*)$, and is therefore
not a parameter of $O(\lqcd)$.

The divergent parts of the counterterms, either $\delta \overline{ m}$
or $\delta \overline{ m}(t^*)$, are gauge invariant, in spite of the
fact that they are calculated from the heavy quark propagator in a
fixed gauge. The argument goes as follows. The linear divergence is
eliminated from any correlation function, i.e. for any external state,
by subtracting from the action a term proportional to the
gauge-invariant operator $\bar h h$. Since in this way one eliminates
all divergences both for quark and hadron external states, the
coefficient of the mixing has to be gauge-invariant. 
This must be true also for the finite non-perturbative contributions
which arise from the coefficient of the linear divergence. These 
contributions are 
generated by terms  which are exponentially small in the
inverse coupling constant, see ref.~\cite{ms,mms}.
As for the finite
parts, the gauge dependence, when it is present, is eliminated from
physical quantities up to the order in perturbation theory at which
the calculation is performed, see e.g.\ the case of the running mass
discussed below.

\par 
In order to obtain the renormalized $\labar$, in addition to the mass
counter-term $\delta \overline{m}$, we need to compute, the ``bare"
binding energy ${\cal E}$.  In the lattice HQET ${\cal E}$, which
diverges linearly in $1/a$, is computed from the two-point correlation
function
\be
 C(t)= \sum_{\vec x}\, \langle 0|\, J(\vec x,t)
J^\dagger(\vec 0,0)\,|0\rangle \, ,
\label{eq:cdef}
\ee where $J$ ($J^\dagger$), is any interpolating operator which can
annihilate (create) a $B$-meson~\footnote{In this paper we only
  consider pseudoscalar B-mesons, but an analogous discussion holds
  for other beauty hadrons, such as the $\Lambda_b$.}.  For a
sufficiently large Euclidean time $t$,
\be
C(t)\rightarrow Z^2 \exp(-{\cal E}t)
\label{eq:ctbig}
\ee
where $Z$ is a constant. 

\par 
We now define the renormalised binding energy by
\be
\labar\, \equiv\, {\cal E}\, -\, \delta \overline{m}\, ,
\label{lare}
\ee
and a ``subtracted'' pole mass by 
\be
m_b\, \equiv\, m_B\, -\, {\cal E}\, +\, \delta \overline{m}\ .
\label{lare2}
\ee
$\lb$ and $m_b$ defined in this way are free of both renormalon
ambiguities and linear divergences.
Similar formulae hold in the case when $\delta \overline{m}(t^*)$ is
used instead to define $\lb(t^*)$ and $m_b(t^*)$.

\section{Numerical implementation of the renormalisation procedure} 
\label{numerical}
As explained in the previous section, the determination of $\lb$
 requires the computation of the heavy  quark propagator 
in a fixed gauge (in order to
obtain the subtracted operator), as well as the evaluation of 
the two-point correlation function (\ref{eq:cdef}).
We have obtained our results using
three independent numerical simulations, whose main parameters
are given in table~\ref{tab:sets}.
\begin{table} \centering
\begin{tabular}{||c|c|c|c||}
\hline
\hline
simulation & volume & $\beta$& Number of configurations
\\ \hline  \hline
 \underline{set A}& $24^{3}\times 40$ &$6.0$&
$500$ \\ \hline 
  \underline{set B}& $24^{3}\times 40 $& $6.2$&
$300$  \\ \hline
 \underline{set C} &$24^{3}\times 40$ &$6.4$&
$300$ \\ \hline \hline
\end{tabular}
\caption{\it{Parameters of the numerical simulations, the
 results of which have been used for the present study.}}
\label{tab:sets}
\end{table}
\par 
Our best value of the subtracted binding energy,
$\bar \Lambda= {\cal E} - \delta \overline{m}$, has been determined by
combining the values of $\delta \overline{m}$ obtained using
\underline{set A}, \underline{set B} and \underline{set C} with the calculation
of ${\cal E}$ performed by the APE collaboration at $\beta=6.0$,
$6.2$ \cite{apeallton} and $6.4$ which will be published elsewhere 
\cite{apeallton2}.
 $\cal E$
was  determined using the SW-Clover fermion action 
\cite{clover} for the light quarks
in the quenched approximation. The
calculations were performed at several values of
the masses of the light quarks, so that
extrapolations to the chiral limit are possible.
$\delta \overline{m}$ ($\delta \overline{m}(t^*)$) has been determined 
by using either the standard definition of the lattice heavy-quark propagator
or its improved (up to and including $O(a)$ terms) version, the explicit
espression of which can be found in ref.~\cite{cgms}. Only the results
obtained by using the improved heavy-quark propagator are reported in the
following. 
\par
All the errors have been computed  with the jacknife method by
decimating  10 
 configurations at a time for \underline{set A}, and 6 configurations at
a time for   \underline{set B} and \underline{set C}.
The error on ${\cal E}$ was
computed with the jacknife method also and we refer the reader to 
refs.~\cite{apeallton,apeallton2} for details.   

\subsection{Determination of the residual mass $\delta \overline{m}$}
\label{mbmsb} 
 
To determine the residual mass, we have to compute the effective mass
\be
a \,  \delta \overline{m}(t)\, 
=\, -\, \ln\left(\frac{S_{H}(t+a)} {S_{H}(t)}\right) 
\label{effmas}
\ee
of the heavy quark propagator, $S_{H}(t)$, defined as
\be
S_{H}(t)\, =\, \frac{1}{3 V}\, \sum_{\vec{x}}\, 
\langle {\mathrm Tr}\left[\, S(t\, , \vec x )\, \right]
\rangle ,\ee
where the trace
is over the colour indices and $V$ denotes the spatial volume of our lattice. 
It is this averaged propagator $S_{H}(t)$, which has been used in the 
computations below; the averaging over all spatial points at time $t$
reduces the statistical errors enormously.

\begin{figure}
\vspace{9pt}
\begin{center}\setlength{\unitlength}{1mm}
\begin{picture}(160,100)
\put(0,-40){\includegraphics{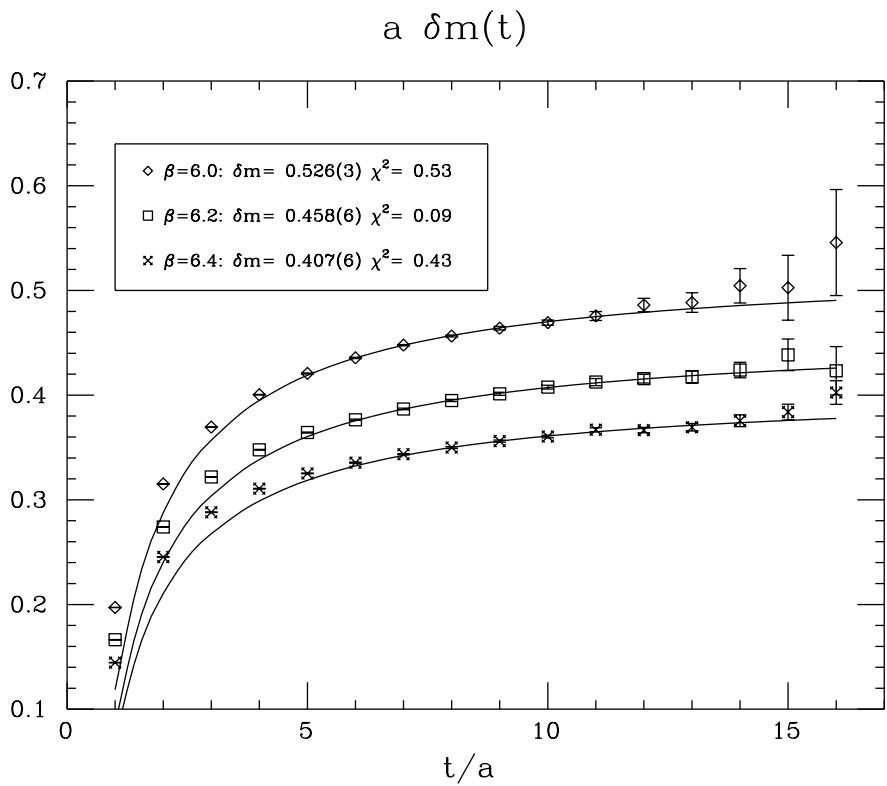}}
\end{picture}
\end{center}
\caption{\it{Effective mass of the improved heavy-quark propagator $S_{H}(t)$,
obtained at $\beta=6.0$ using $500$ configurations, and at $\beta=6.2$ and
$6.4$ using 300 configurations, as a 
function of the time. The curves represent   fits of the numerical results 
to the expression given in eq.~(12).}}
\label{fig:fit60}
\end{figure}
In fig.~\ref{fig:fit60}  we present the values of
$\delta \overline{m}(t)$ for the
improved  propagator as a function of $t/a$. The figure shows the very
high accuracy which can be reached in the determination of the effective
mass, i.e. in the determination of the logarithmic derivative (\ref{ct})
of the heavy quark propagator. This will allow us to make a careful 
study of the infrared behaviour of the quark propagator.
\par As shown in ref.~\cite{cgms},
the effective mass is indistinguishable in the improved and unimproved
cases, for $t/a > 4$--$5$. Thus, in order to minimize lattice artefacts,
 we have only used  the results obtained for $t/a \ge 6$.
\par  Inspired 
by  the results of one-loop perturbation theory \cite{ms},
 we made a fit to $\delta \overline{m}(t)$ using the expression 
\be a \, \delta \overline{m}(t)  = a \, \delta \overline{m} + 
\gamma\frac{a}{t}\ . 
\label{eq:fit}
\ee
In order to mimic higher order effects, we have also used different
expressions to fit $\delta \overline{m}(t)$, e.g.
\be a \, \delta \overline{m}(t)  =a \,  \delta \overline{m} + \gamma^\prime
\ln \Bigl( \frac{t+a}{t}\Bigr)  \label{eq:fit1}
\ee
or 
\be a \,  \delta \overline{m}(t)  = a\, \delta \overline{m} - 
{\gamma^{\prime\prime}}\ln\Bigl(\frac{\alpha_s[{\cal K}/(t+a)]}
{\alpha_s[ {\cal K}/t]}
\Bigl) \to
a \, \delta \overline{m} +
\gamma^{\prime\prime}\ln\Bigl(\frac{\ln[(t+a)]+{\cal C}}{\ln[t]+ {\cal C}}
\Bigl),  \label{eq:fit2}
\ee
and changed the interval of the fits in order to check the stability
of the determination of $ \delta \overline{m}$. 
 Finally, in order to monitor possible confining effects in the 
heavy-quark propagator, we have also tried a fit of the form
\be a \, \delta \overline{m}(t)  =a \,  \delta \overline{m} + 
\widetilde \gamma
\ln \Bigl( \frac{t+a}{t}\Bigr)  + \rho t\, ,\label{eq:fitcon}
\ee
where the last term in eq.~(\ref{eq:fitcon}) will be discussed
later on.
In eqs.~(\ref{eq:fit})--(\ref{eq:fitcon}),
$\delta \overline{m}$, $\gamma, \dots, \gamma^{\prime\prime}$, ${\cal C}$
and $\rho$ are free parameters of the fit. 
\par From the different results obtained by varying the
 fitting functions (\ref{eq:fit})--(\ref{eq:fit2}) and the
time intervals,  we quote
\be a \,\delta \overline{m} = 0.526 \pm 0.003 \pm 0.006  \hbox{ at }
 \beta=6.0 \label{dbm60} \ee
\be a \,\delta \overline{m} = 0.458 \pm 0.005 \pm 0.004 \hbox{ at }
 \beta=6.2 \label{dbm62} \ee
\be a \,\delta \overline{m} = 0.407 \pm 0.006 \pm 0.006 \hbox{ at }
 \beta=6.4 \label{dbm64} \ee
where in all  cases the first error is statistical, and the second is
an estimate of the systematic uncertainty, based on the spread of results
obtained using different time intervals and fitting functions. 
Notice the improvement of the above
figures, when compared to our old results $a \,\delta
 \overline{m} = 0.521 \pm 0.006 \pm 0.010$ at $\beta=6.0$
and $a \,\delta \overline{m} = 0.445 \pm 0.008 \pm 0.010$ at $\beta=6.2$.
The improvement is due to the increased statistics, which also allowed
us to study the heavy quark propagator at larger time distances, thus
reducing also the systematic error.
\par In order to obtain $\bar \Lambda$ we have used:
\begin{itemize} 
\item $\delta \overline{m}$ from eqs.~(\ref{dbm60})--(\ref{dbm64});
\item  the SW-Clover determination of ${\cal E}$ of the APE collaboration,
$a {\cal E}=0.610 \pm 0.010$ at $\beta=6.0$,
 $a {\cal E}=0.520 \pm 0.010$ at $\beta=6.2$ \cite{apeallton}
and $a {\cal E}=0.460 \pm 0.007$ at $\beta=6.4$~\cite{apeallton2}; 
\item $a^{-1}(\beta=6.0)=2.0 \pm 0.2$~GeV,  
$a^{-1}(\beta=6.2)=2.9 \pm 0.3$~GeV
and   $a^{-1}(\beta=6.4)=3.8 \pm~0.3$ GeV.
A significant contribution to the final error in $\labar$ comes from
the calibration of the lattice spacing in quenched simulations which
typically has an uncertainty of $O(10\%)$, depending on the physical
quantity which is used to set the scale. We take the results given
above as a fair representation of the spread of values of $a^{-1}$.
\end{itemize} 
We then find
\beqn
\lb & = &{\cal E}-\delta \overline{m} \,=\,
170 \pm 30\ {\mathrm MeV}\ \mbox{at}\ \beta =6.0 
\label{eq:lb6.0}\\ 
\lb & = & {\cal E}-\delta \overline{m} \,=\,
180 \pm 40\ {\mathrm MeV}\ \mbox{at}\ \beta =6.2 
\label{eq:lb6.2}
\\ 
\lb & = & {\cal E}-\delta \overline{m} \,=\,
200 \pm 40\ {\mathrm MeV}\ \mbox{at}\ \beta =6.4 
\label{eq:lb6.4}\eeqn
where the statistical errors have been combined in quadrature with
those due to the uncertainty in the lattice spacing. 
\par
Within the uncertainties, the results in
eqs.~(\ref{eq:lb6.0})--(\ref{eq:lb6.4}) are compatible with the
expected independence of $\bar\Lambda$ of the lattice spacing. Given
the intrinsic uncertainty in the value of the lattice spacing in
quenched simulations it is difficult however, to check this more
precisely. Certainly our results exclude large $O(a)$ effects.
Assuming that $\bar \Lambda$ is indeed constant, we combined together
the results of eqs.~(\ref{eq:lb6.0})--(\ref{eq:lb6.4}) to obtain 
\be
\bar \Lambda= (180 ^{+30}_{-20} ) \hbox{ MeV }\, ,
\label{lbres1} \ee
where in the final error we have included also the discretization
error which has been conservatively estimated to be of the order of
$+20$ MeV in ref.~\cite{cgms}. The estimate of the discretization
error was obtained by comparing the value of ${\cal E}$ obtained by
using the standard Wilson action and the SW-Clover action at the same
values of $\beta$.

It would be interesting to compare this value of $\labar$ with other
determinations, obtained by using different methods, e.g.\ the recent
result obtained from the semileptonic width in ref.~\cite{kapustin}.
The relation between our definition and other ones, provided they are
implemented consistently and are free of renormalon ambiguities, can
be formally derived in perturbation theory. Even in the case where the
perturbative series which connects two different definitions of
$\labar$ is precisely determined, this would not reduce the error in
the prediction for physical quantities, such as $\overline{m}_b$.
Indeed, as discussed below, for $\overline{m}_b$ the error of about
200~MeV is dominated by the perturbative series which matches QCD to
the effective theory, and not by the error in $\labar$.

\begin{table} \centering
\begin{tabular}{||c|c|c|c|c||}
\hline
\hline
{$\beta$} &
{$t^*/a$}&$a\delta\overline{m}(t^*)$&{$\pi/t^*$ (GeV)}
&{$\bar \Lambda(t^*)$ (MeV)}\\
\hline
\hline
$6.0$ &$3$& $0.3694(7)$ &$2.1 \pm 0.2$ &$ 480 \pm 20 \pm 50$ \\ 
$$ &$4$   & $0.4004(8)$&$1.6 \pm 0.2$ &$420 \pm 20 \pm 40$ \\
$$ &$5$   & $0.4207(8)$&$1.3 \pm 0.1$ &$380 \pm 20 \pm 40$ \\
$$ &$6$   & $0.4356(8)$&$1.0 \pm 0.1$ &$350 \pm 20 \pm 40$ \\
\hline
$6.2$&$4$ &$0.3478(6)$ &$2.3 \pm 0.2$ &$ 500 \pm 20\pm 50$ \\
$$&$5$ & $0.3643(6)$ &$1.8 \pm 0.2$ &$ 450 \pm 20 \pm 50$ \\
$$&$6$ & $0.3765(10)$  &$1.5 \pm 0.2$ &$ 420 \pm 20 \pm 40$ \\
$$&$7$ & $0.3868(12)$  &$1.3 \pm 0.1$ &$ 390 \pm 20 \pm 40$ \\
$$&$8$ & $0.3949(15)$  &$1.1 \pm 0.1$ &$ 360 \pm 20 \pm 40$ \\
\hline
$6.4$&$5$ &$0.3252(4)$ &$2.4 \pm 0.2$ &$ 510 \pm 30\pm 40$ \\
$$&$6$ & $0.3355(4)$ &$2.0 \pm 0.2$ &$ 470 \pm 30 \pm 40$ \\
$$&$7$ & $0.3435(5)$  &$1.7 \pm 0.1$ &$ 440 \pm 30 \pm 30$ \\
$$&$8$ & $0.3499(7)$  &$1.5 \pm 0.1$ &$ 420 \pm 30 \pm 30$ \\
$$&$9$ & $0.3561(7)$  &$1.3 \pm 0.1$ &$ 390 \pm 30 \pm 30$ \\
$$&$10$ & $0.3604(13)$  &$1.2 \pm 0.1$ &$ 380 \pm 30 \pm 30$ \\
$$&$11$ & $0.3669(18)$  &$1.1 \pm 0.1$ &$ 350 \pm 30 \pm 30$ \\
\hline
\hline
\end{tabular}
\caption{\it{Results for $\bar \Lambda(t^*)={\cal E}-\delta
\overline{m}(t^*)$ for different normalization times $t^*$,
using the results from \underline{set A}--\underline{set C}.
The first error on $\bar \Lambda(t^*)$ is obtained by combining
 the errors on ${\cal E}$ and $\delta
\overline{m}(t^*)$ 
in quadrature; the second error (and the error on $\pi/t^*$)
comes from the calibration of the lattice spacing.}}
\label{tab:lt}
\end{table}
\par The determination of the mass counter-term $
\delta \overline{m}(t^*)$ at fixed $t=t^*$, requires
no fitting, and the results obtained using \underline{set A}--\underline{set C}
are presented in table~\ref{tab:lt}. Only the results
obtained in a range of $t^*$ corresponding to  values of $\pi/t^*$ between
$1.0$ and $2.5$ GeV are reported in the table. The results for 
$\labar(t^*)$ obtained using the same values of ${\cal E}$ and
of $a^{-1}$ as before, are also reported in the same table. They
show that  the  values of the binding energy  
at the same value of the renormalisation scale $\pi/t^*$ (but at different
values of the lattice spacing) are very consistent, implying  that lattice
artefacts are rather small.

\par We conclude this subsection with a brief discussion of
the infrared behaviour of the heavy quark propagator.  Due to
confinement effects, it is not clear that the logarithmic derivative
(\ref{effmas}) has a finite limit as $t \to \infty$.  Indeed, if
(\ref{effmas}) goes to a constant at large time distances, then the
propagator in the effective theory behaves like an on-shell propagator
of a particle of mass $\delta \overline{m}$, $S_H(t) \sim \exp(-\delta
\overline{m}\, t)$.  This contradicts the n\"aive expectation that the
Fourier transform of the propagator of a confined object should not
have poles, as was found, for example, in lattice computations of the
gluon propagator \cite{bsg}. In the case of the heavy-quark
propagator, it is also possible that the ``effective" mass $\delta
\overline{m}(t)$ does not tend to a constant at large time-distances,
and that the propagator decreases faster than an exponential. In order
to monitor this, we have allowed for a linear term in $t$ in the
effective mass and have fitted (\ref{effmas}) to (\ref{eq:fitcon}).
We found that the slope $\rho$ is compatible with zero \footnote{ For
  example $\rho=0.002 \pm 0.002$ at $\beta=6.0$ by fitting in the
  interval $6 \le t/a \le 16$.} for any fitting interval, and at all
the values of $\beta$ considered in this study.  Morevover the value
$\delta \overline{m}$ obtained by fitting $\delta \overline{m}(t)$ to
eq.~(\ref{eq:fit1}) or to eq.~(\ref{eq:fitcon}) are equal within the
errors.  This implies that no sizeable variation of the effective
mass, due to confinement effects, is observed down to rather low scales
(as small as $\pi/16 a \sim 390$ MeV, or distances up to 1.6\,fm, at
$\beta=6.0$), which belong to the non-perturbative region.  

\subsection{The $\overline{MS}$ mass of the $b$-quark}
\label{subsec:mbar}

Although it is convenient and conventional to introduce parameters
such as $\lb$ when studying the $O(1/m_b)$ corrections to quantities
in heavy quark physics, it is not necessary. One can instead derive
physical quantities from results obtained using lattice simulations.
In this subsection we present the results for the $\overline {MS}$
mass of the $b$--quark, from simulations in the Heavy Quark Effective
Theory~\cite{cgms}. The evaluation of a short distance mass does not
require any assumptions about the long-distance behaviour of the quark
propagator, indeed it can be obtained from computations of the meson
propagator only. In ref.~\cite{cgms} it was shown that, at one-loop order,
\beq 
\overline{m}_b=\Bigl(M_B-{\cal E}+ \alpha_s(a)\frac{X}{a}\Bigr)
 \Bigl(
1-\frac{4\alpha_s(\overline {m}_b)}{3 \pi} \Bigr)\ ,
\label{matchid} \eeq
where $X$ is the coefficient of the one-loop mass term, obtained in
the lattice HQET with zero bare mass. The second factor on the
right-hand side of eq.~(\ref{matchid}) is that which relates the pole
mass to $\overline{m}_b$. The perturbative series of linearly divergent
terms (of which only the one-loop term is explicitly exhibited in
eq.~(\ref{matchid})) removes the linear divergence form ${\cal E}$, but
has a renormalon ambiguity of $O(\lqcd)$. This ambiguity is cancelled
by that in the second factor, i.e the series relating $\overline{m}_b$
to the pole mass. Thus $\overline{m}_b$ has neither a linear
divergence not a renormalon ambiguity. In evaluating $\overline{m}_b$
we take the experimental value of $M_B$ (5.278 GeV) and the value of
${\cal E}$ as measured in the lattice simulations of the APE
collaboration. $\alpha_s(a)$ is taken in the range $0.13$ and $0.18$.

We also evaluate $\overline{m}_b$ using 
\beq 
\overline{m}_b=m_b(t^*) \times \overline{C}_m(t^*)\, ,
\label{matchi}\eeq
where 
\beq m_b(t^*)\equiv M_B-\bar \Lambda(t^*)=M_B- {\cal E}+ \delta
\overline{m}(t^*)\, , \eeq and 
\beq \overline{C}_m(t^*)\equiv1
-\frac{4\alpha_s( \overline {m}_b)}{3 \pi} - \frac{1}{\overline{m}_b}
\left( \delta \overline{m}(t^*) -\alpha_s(a) \frac{X}{a} \right) \,.
\eeq 
Eq.~(\ref{matchi}) is equivalent to (\ref{matchid}) up to one-loop
order and neglecting terms of $O(1/m_b)$ in the quark mass.
Numerical differences in the results
obtained in these two ways are a partial measure of the systematic
uncertainties.  The values of $a \, \delta \overline{m}(t^*)$ used in
the analysis are taken from eqs.~(\ref{dbm60})---(\ref{dbm64}) and
table~\ref{tab:lt}.

To obtain a distribution of results, we varied ${\cal E}$,
$\Lambda_{{\mathrm QCD}}$ and $a \, \delta \overline{m}(t^*)$
according to a gaussian distribution; $a^{-1}$ was varied with
flat distribution within its error, while $\alpha_s(a)$ was written
in terms of the leading quenched expression of the running
coupling constant, evaluated at
the scale $\pi/a$, with $\Lambda_{{\mathrm QCD}}$ distributed according to
a flat distribution of width $\sigma$
 and such that $\alpha_s=0.13$ for $\Lambda_{{\mathrm QCD}}-\sigma$
and $\alpha_s=0.18$ for $\Lambda_{{\mathrm QCD}}+\sigma$. 
The resulting distribution is a pseudo-gaussian \cite{cgms}.
From the width
of the distribution we estimate the average value and  error on 
$\overline {m}_b$. The results obtained using eqs.~(\ref{matchi})
and (\ref{matchid}) are given in table~\ref{tab:mbar}.
\begin{table}[h] \label{tmass}\centering
\begin{tabular}{||c|r|r|r||r||}
\hline
\hline
\multicolumn{5}{||c||}{$\overline{MS}$ running $b$-quark mass, 
$\overline{m}_{b}(\overline{m}_{b})$}\\ 
\hline
\hline
\multicolumn{1}{||c|}{$\beta$}&\multicolumn{1}{c|}{Eq. (24) $t^{*} = 3-6$}&
\multicolumn{1}{c|}{Eq. (24) $t^{*} = \infty$}&\multicolumn{1}{|c||}{Eq. (22)}&
\multicolumn{1}{c||}{$\overline{m}_{b}(\overline{m}_{b})$ (GeV)}\\
\hline
\hline
$6.0$&$4.20(7)$&$4.18(7)$&$4.22(7)$&$4.20(7)$\\
$6.2$&$4.13(9)$&$4.10(9)$&$4.15(8)$&$4.13(9)$\\
$6.4$&$4.07(9)$&$4.04(9)$&$4.09(8)$&$4.07(9)$\\
\hline
\hline
\multicolumn{4}{||c||}{AVERAGE}&\multicolumn{1}{c||}{$4.15\, \pm\, 0.05\, \pm 0.05$}\\
\hline
\hline
\end{tabular}
\caption{The $\overline{MS}$ running mass of the $b$ quark.
The first error is statistical and the second systematic.}
\label{tab:mbar}
\end{table}

From the results in the table, we estimate $\overline {m}_b=4.15 \pm
0.05 \pm 0.05 \hbox{ GeV}$, where the second error is the systematic
error, evaluated by comparing the results obtained using the different
formulae given in eqs.~(\ref{matchi}) and (\ref{matchid}). This error,
which is of about $50$~MeV, can be interpreted as being due to higher
order corrections in $\alpha_s$. It can be compared with the estimate
of the effects of higher order corrections, derived from the the
bubble-resummation approximation, which is about $200$ MeV
\cite{emphasis}. We believe that an estimate of $200$ MeV is a more
realistic one \footnote{The ignorance of higher order corrections also
  implies an error of this order in any approach.} and for this reason
we quote as our final result 
\beqn \overline {m}_b=4.15 \pm 0.05 \pm
0.20 \hbox{ GeV}\, . 
\eeqn

\section{Conclusions} \label{conclu}

In this study we have presented a high statistics
calculation of the binding energy of the $b$-quark in a $B$-meson.
The improved statistics, with respect to ref.~\cite{cgms}
has allowed a better control of the infrared behaviour of the
heavy quark propagator. There is no point in further increasing 
the statistics in the calculation of $\delta \overline{m}$, 
since the precision on $\labar$
is mainly  limited by the calibration of the lattice spacing
in quenched simulations and by the precision in the determination
of the bare binding energy ${\cal E}$. It is difficult to
imagine a reduction of the uncertainty
in the calibration of the lattice spacing in the quenched case,
while  the error on the bare
 binding energy is likely  to be reduced in the near
future.

\par The main error in the evaluation of the 
$\overline{MS}$ $b$-quark mass ($\sim 200$ MeV) comes from the unknown
higher order terms in the perturbative matching of the HQET to the
full theory \cite{emphasis}.  This error can be reduced by computing
perturbative corrections at two-loop order.  This requires the
calculation of the heavy quark propagator in the continuum and of the
propagator in the lattice effective theory at $O(\alpha_s^2)$. The
continuum calculation has already been performed \cite{broad}. We
believe that the best approach to the two-loop (and higher order)
calculations in the lattice effective theory is the one based on the
Langevin equation being developed by the authors of
ref.~\cite{marche}. Based on the summation of light-quark bubble
graphs \cite{emphasis,bbb}, we estimate that knowledge of the matching
factor at two-loops would reduce the theoretical uncertainty to about
100~MeV.

\section*{Acknowlegments} 
We acknowledge the partial support by the EC contract 
CHRX-CT92-0051.
V.G. acknowledges the partial support by CICYT under grant number 
AEN-96/1718.    
G.M. acknowledges the partial support by M.U.R.S.T. 
C.T.S. acknowledges the Particle Physics and Astronomy Research Council
for its support through the award of a Senior Fellowship.

\end{document}